\documentstyle[11pt,twoside,pasp3D,epsf]{article}

\def\arcdeg{\hbox{$^\circ$}}

\def\arcsec{\hbox{$^{\prime\prime}$}}

\begin{document}

\title{Fabry-Perot Imaging Spectroscopy of Starburst and AGN Winds}

\author{S. Veilleux}

\affil{Department of Astronomy, University of Maryland, College Park, MD 20742}

\begin{abstract}

To date, the most detailed studies of galactic winds have come from
3-D spectrophotometric observations with radio and Fabry-Perot
interferometers. Here, we report the latest results from a long-term
optical survey of nearby active and starburst galaxies with the Hawaii
Imaging Fabry-Perot Interferometer (HIFI) at Mauna Kea and the
TAURUS-2 system in Australia. These data reveal that the outflows are
highly complex, highly energetic ($>$ 10$^{53}$ ergs in most cases),
and the brightest emission often appears to be associated with strong
shocks. The outflowing material in the starburst galaxies generally
lies on the surface of bubbles or along the walls of funnel-shaped
winds rapidly accelerating out of the galactic plane.  These winds are
sometimes lop-sided and tilted with respect to the polar axis of the
host galaxy. Evidence for entrainment of (rotating) disk material is
seen in some objects.  Our results are combined with HST, radio and
X-ray data and discussed in the context of future surveys of distant
galaxies on 8-meter class telescopes. 

\end{abstract}

\section{Introduction}

Active galactic nuclei (AGN) and nuclear starbursts may severely
disrupt the gas phase of galaxies through deposition of a large amount
of mechanical energy in the centers of galaxies.  As a result, a
large-scale galactic wind (``superwind'') that encompasses much of the
central regions of these galaxies may be created (e.g., Chevalier \&
Clegg 1985; Schiano 1985).  Depending upon the extent of the halo and
its density and upon the wind's mechanical luminosity and duration,
the wind may ultimately blow out through the halo and into the
intergalactic medium.  The effects of these superwinds may be
far-reaching. Bregman (1978) has suggested that the Hubble sequence
can be understood in terms of a galaxy's greater ability to sustain
winds with increasing bulge-to-disk ratio. Superwinds may affect the
thermal and chemical evolution of galaxies by depositing large
quantities of hot, metal-enriched material on the outskirts of
galaxies. They also offer a natural way to create a cosmically
evolving population of large, metal-enriched, kinematically-complex
gaseous halos, in many ways resembling the sharp metal lines and
Lyman-limit systems observed in quasar spectra.

Strong evidence for spatially-resolved superwinds now exists in
several nearby starburst, Seyfert, and dwarf galaxies (e.g., Bland \&
Tully 1988; Cecil et al. 1990; Heckman et al. 1990; Lehnert \& Heckman
1996; Marlowe et al. 1995; Meurer et al. 1992; Veilleux et al.  1994;
Colbert et al. 1996).  Our group is combining Fabry-Perot imaging
spectrophotometry with radio and X-ray data to track the energy flow
through various gas phases. The complete spatial and kinematic
sampling of the Fabry-Perot (FP) data is ideally suited to study the
complex and extended morphology of the warm line-emitting material
that is entrained in the wind flow. The radio and X-ray data
complement the FP data by probing the relativistic and hot gas
components, respectively.  The high level of sophistication of recent
hydrodynamical simulations (e.g., Tomisaka 1990; Slavin \& Cox 1992;
Mineshige et al. 1993; Suchkov et al. 1994) has provided the
theoretical basis to interpret our data and to predict the evolution
and eventual resting place (disk, halo, or intergalactic medium) of
the outflowing material. In the present paper, we summarize the
results on three representative objects from our Fabry-Perot survey
and discuss possible implications.

\section{The Fabry-Perot Survey}

Using the gap-scanning mode of the Hawaii Imaging Fabry-Perot
Interferometer (HIFI) on Mauna Kea Observatories and of the TAURUS-2
Fabry-Perot system on the Anglo-Australian Telescope, our group is
carrying out a survey of twenty nearby ($z$ $<$ 0.01) starburst and
Seyfert disk galaxies. Deep tunable-filter images and stare-mode
FP spectra supplement some of the data cubes. The objects in our sample
were selected on the basis of {\em a priori} evidence for large-scale
nuclear outflows.

So far, high-quality data cubes have been obtained and analyzed for
about a dozen galaxies, and the results have been published for ten
of them. Our FP data set allows us to perform spectrophotometric
analyses of the line-emitting gas at typically 10,000 -- 100,000
positions across the extent of our sample galaxies. These data
therefore provide very stringent contraints on the general flow
pattern of the line-emitting gas entrained in these outflows.

\section{Results}

Because of space limitations, we focus our discussion of the results
on three of our sample galaxies.  These objects were chosen to
illustrate the broad diversity of morphologies, kinematics, and
energetics associated with galactic winds.

\subsection{M82}

This prototypical starburst galaxy has long been suspected to host a
galactic-scale outflow (e.g., Lynds \& Sandage 1963; Burbidge,
Burbidge, \& Rubin 1964). A recent HST image of this galaxy (Fig. 1)
shows the well-known filamentary complex that extends several kpc
above and below the disk of M82.  The FP data from Shopbell \&
Bland-Hawthorn (1998) reveal a bipolar outflow of material that
originates from the bright starburst regions in the galaxy's inner
disk but is misaligned with respect to the galaxy spin axis. The
deprojected outflow velocity indicated by the optical filaments
increases with radius from 525 to 655 km s$^{-1}$. Double components
are detected in the centers of the outflowing lobes, with line
splitting by $\sim$ 300 km s$^{-1}$ over a region almost 1 kpc in
size. The lobes lie along an axis tilted by 15$\arcdeg$ with respect
to the spin axis of the galaxy.

\begin{figure}
\vskip 4.0in
\caption{Continuum-subtracted H$\alpha$+[N~II] HST/WFPC2 image of M82
showing dramatic filamentation and bow shock structures in the
outflowing lobes. The size of the smallest resolved features in this
image is about 2 pc (Shopbell et al. 1999)}

\end{figure}

The filaments are not simple surfaces of revolution, nor is the
emission distributed evenly over the surfaces. These lobes are best
modeled as a composite of cylindrical and conical structures,
collimated in the inner $\sim$ 500 pc but expanding at a larger
opening angle of $\sim$ 25$\arcdeg$ beyond that radius. 
The wind in M82 therefore seems to be freely flowing into the galaxy
halo (``free-wind'' phase in the nomenclature of Weaver et al. 1977). 
Using this outflow geometry and assuming a filling factor of
0.1, Shopbell \& Bland-Hawthorn (1998) finds that a kinetic energy of
$\sim$ 2 $\times$ 10$^{55}$ is involved in the outflow. There is also
some evidence for rotation of the wind filaments about the outflow
axis in support of entrainment.

The observed filamentation probably arises from large-scale
shocks from the high-speed wind plowing into the gaseous halo and
entrained disk material. The line ratios suggest that photoionization
by the nuclear starburst play a significant role in the excitation of
the optical filament gas, but that shock ionization becomes
increasingly important at large radii.

\subsection{NGC 3079}

In no other galaxy is the impact of superwinds more evident than in
NGC~3079, a nearby edge-on spiral galaxy which is host to a
spectacular kpc-scale bubble (see Fig. 2).  Violent gas motions that
range over 2,000 km s$^{-1}$ are detected across the bubble and
diametrically opposite on the other side of the nucleus. The unusual
gaseous excitation (e.g., [N~II] $\lambda$6583/H$\alpha>$ 1) of the
line-emitting gas in the bubble region confirms that shocks are
important.

\begin{figure}
\vskip 3.4in

\caption{(left panel) Continuum-subtracted H$\alpha$+[N~II]
HST/WFPC2 image of the nuclear bubble in NGC 3079. The spatial
resolution is about 8 pc. (upper right panels) Results of numerical
simulations from Suchkov et al. (1994). Note the resemblance with the
fine structures observed in the HST image. (lower right panel)
A view of the star-forming disk in NGC 3079 derived from the
HST images. Elevated line-emitting chimneys and filaments are seen
near the brightest H~II regions. (Cecil et al. 1999)}

\end{figure}

This is the most powerful example known of a windblown bubble ($\sim$
2 $\times$ 10$^{56}$ N$_e^{-1}$), and an excellent laboratory to study
wind dynamics. An ovoidal bubble, inflated from the nucleus with
monotonically increasing velocities (V $\propto$ R$^n$ with 2 $< n <$
3) and inclined $\sim$3$\arcdeg$ from the plane of the sky, provides a
good first-order fit to the observed velocity field.  The dimensions
and energies of the bubble imply that it is in the blowout phase and
partially ruptured. A detailed dynamical analysis of this outflow
indicates that the wind alone can contribute up to 5$t_{\rm
outflow,8}$\% of the total metal content of the host galaxy, where
$t_{\rm outflow,8}$ is the outflow lifetime in units of 10$^8$ yr
(Veilleux et al. 1994).

The core of NGC 3079 harbors both a nuclear starburst and an AGN. The
nature of the energy source that drives the outflow is not clear at
present. The poorly constrained electron density and filling factor of
the warm line-emitting material in the bubble are the main sources of
uncertainty in the calculation of the energetics.  To shed new light
on this issue, we recently obtained a WFPC2 HST image of NGC 3079 with
resolution of $\sim$ 0$\farcs$1 ($\sim$ 8 pc; Fig 2). The HST image
reveals intricate patterns in the line-emitting material near the top
of the bubble. These features share a remarkable resemblance with
those observed in hydrodynamical simulations (e.g., Suchkov et
al. 1994) -- they are probably the signatures of Rayleigh-Taylor
instabilities in the entrained material.  Similar filamentary
structures are observed near the brightest star-forming regions in the
disk of NGC~3079, bringing further support to the idea of a
dynamically active disk in this galaxy (Veilleux et al. 1995).  A
preliminary quantitative analysis of the HST data suggests that the
AGN in NGC~3079 is not contributing significantly to the nuclear
outflow.

\subsection{Circinus}

The FP data on Circinus show a complex of ionized filaments extending
radially from the nucleus out to distances of 1 kpc (Fig. 3). The
velocity field of the filaments confirms that they represent material
expelled from the nucleus or entrained in a wide-angle wind roughly
aligned with the polar axis of the galaxy.  Extrapolation of these
filaments to smaller radii comes to within 1$\arcsec$ of the active
galactic nucleus, therefore suggesting a AGN or nuclear starburst
origin to these features.  The outflow involves a fairly modest 
kinetic energy ($\sim$ 10$^{53}$ N$_{e,2}^{-1}$ ergs) and therefore
appears to lie at the low energy end of the distribution for wide-angle
events observed in nearby galaxies.

The complex of radial filaments and bow shocks detected in the
Circinus galaxy is unique among active galaxies and does not fit
within the standard evolutionary picture of windblown bubbles (e.g.,
Weaver et al. 1977).  It is not clear at present why
that is the case.  The discovery of these features in the Circinus
galaxy, a spiral galaxy with an abnormal richness of gas (Freeman et
al. 1977), brings up the possibility that we may be witnessing a
common evolutionary phase in the lives of gas-rich active galaxies
during which the dusty cocoon surrounding the nucleus is expelled by
the combined action of jet and wind phenomena.

\begin{figure}
\vskip 3.1in
\caption{Line flux images of Circinus, the nearest Seyfert galaxy:
$a$, [O~III] $\lambda$5007 and $b$, blueshifted H$\alpha$.  The
position of the infrared nucleus is indicated in each image by a
cross. The spatial scale, indicated by a horizontal bar at the bottom
of the [O~III] image, corresponds to $\sim$ 25 arcsec~or 500~pc.  Note
the unusual complex of radial filaments emerging from the
nucleus. PA$_{\rm major}$(disk) = 30$^\circ$, $i$(disk) = 65$^\circ$
(Veilleux \& Bland-Hawthorn 1997)}
\end{figure}

\section{Future Avenues of Research}

The overall agreement between the simulations of Suchkov et al. (1994)
and current observations of local galaxies with galactic winds augurs
well for the future.  However, substantial quantitative differences
still remain: e.g., current models severely underestimate the outflow
velocities of the entrained line-emitting material (by more than an
order of magnitude in NGC 3079).  A more realistic treatment of the
multi-phase ISM in the host galaxy may help solve some of these
problems and may help explain the morphological peculiarities of the
outflow in the gas-rich galaxy Circinus.

The results from these observational and theoretical studies will
eventually serve as a critical local baseline for future deeper
surveys with IFUs on 8-meter class telescopes from the ground and in
space. Current surveys suggest that galaxies have experienced a very
active phase of star formation and nuclear activity around redshifts
of 1-3 (e.g., Madau, Pozzetti, \& Dickinson 1998; Schmidt, Schneider,
\& Gunn 1995).  Detailed comparisons of high-redshift galaxies with
local superwind hosts and with state-of-the-art numerical simulations
of windblown bubbles will help us quantify the impact starburst- and
AGN-driven winds may have had on the chemical and thermal evolution of
the galactic and intergalactic environments.

\acknowledgments

The survey described in this paper is done in collaboration with
J. Bland-Hawthorn, G. Cecil, P. Shopbell, and R. B. Tully. The author
is grateful for support of this research by a Cottrell Scholarship
awarded by the Research Corporation, NASA/LTSA grant NAG 56547, and
NSF/CAREER grant AST-9874973.

\end{document}